\documentclass [12pt] {report}
\usepackage{amssymb}
\newcommand {\eqdef} {\stackrel{\rm def}{=}}

\newcommand{\arcsinh}{\mathop{\rm arcsinh}\nolimits}
\newcommand {\al} {\alpha}

\newcommand {\ga} {\gamma}
\newcommand {\la} {\lambda}

\newcommand {\de} {\delta}

\newcommand {\fr} {\displaystyle\frac}
\newcommand {\wt} {\widetilde}
\newcommand {\be} {\begin{equation}}
\newcommand {\ee} {\end{equation}}
\newcommand {\ba} {\begin{array}}
\newcommand {\ea} {\end{array}}
\newcommand {\bp} {\begin{picture}}
\newcommand {\ep} {\end{picture}}
\newcommand {\bc} {\begin{center}}
\newcommand {\ec} {\end{center}}
\newcommand {\bt} {\begin{tabular}}
\newcommand {\et} {\end{tabular}}
\newcommand {\lf} {\left}
\newcommand {\rg} {\right}

\newcommand {\cL} {{\cal L}}

\newcommand {\cS} {{\cal S}}

\newcommand {\ses} {\medskip}

\newcommand {\bibit} {\bibitem}
\newcommand {\nin} {\noindent}

\def\2#1#2#3{{#1}_{#2}\hspace{0pt}^{#3}}
\def\3#1#2#3#4{{#1}_{#2}\hspace{0pt}^{#3}\hspace{0pt}_{#4}}

\newcounter{sctn}
\def\sec#1.#2\par{\setcounter{sctn}{#1}\setcounter{equation}{0}
                  \noindent{\bf\boldmath#1.#2}\bigskip\par}

\begin {document}

\begin {titlepage}

\vspace{0.1in}

\begin{center}
{\Large
Nonlinear Relativistic Invariance For Quadrahyperbolic Numbers
}\\
\end{center}

\vspace{0.3in}

\begin{center}

\vspace{.15in}
{\large D.G. Pavlov\\}
\vspace{.25in}
{\it Division of Applied Mathematics, Academy of
Civil Defence.\\
Moscow, Russia\\

pavlovdg@newmail.ru

}

\vspace{.05in}

\end{center}

\begin{abstract}

One may ask whether an extended group of invariance
can naturally be attributed to the space of
associative
commutative Quadrahyperbolic  Numbers?
To search for a rigorous and positive answer to the question,
we shall focus on the method of derivation of
the respective
invariance.
The outcome that there exist 3--parametric nonlinear transformations which
leave invariant  the
scalar product
chosen
appropriately
for
The Quadrahyperbolic  Numbers,
is the main result of the present publication.

\end{abstract}

\end{titlepage}

\vskip 1cm

\setcounter{sctn}{1}
\setcounter{equation}{0}
{\nin\bf 1. INTRODUCTION}
\medskip

In the previous work [1]
we raised several urgent problems which solution is doomed to elucidate
the geometric
as well as
relativistic--geometric
ingredients of
The Space of
Quadrahyperbolic Numbers
(QH--Space for brevity).
In the present paper we make due attempts to explain, from first ideas,
how the group of nonlinear invariance should supplement the linear dilatation
invariance operative trivially
in
the
QH--Space.
As matter stands, to proceed successfully in this direction,
one should propose a convenient scalar product for pairs of
the
Quadrahyperbolic Numbers.
In this respect, the available and attractive possibility is
 to exploit respectively
the
concept of transversality
(introduced in Sec. 4 of [1]).
Namely, it is the symmetrized metric form (1.7)
for pairs of
the
Quadrahyperbolic Numbers
that is a handy and convenient scalar
product.
Adhering to this choice, we prove
that the nonlinear invariance transformations
exist which do
leave our scalar product invariant.
In the dimension
$N=4$
used, the linear unimodular dilatations
involve 3 parameters, and the nonlinear transformations,  -- as
being the transformations which act independently in each of three
basic sectors of the scalar product under study, -- involve also three
parameters in general. Therefore, the fact that
 the total invariance group
is 6--parametrical
is not violated under the transition from the Lorentz invariance of
the pseudoeuclidean space to the proposed invariance of
the
QH-Space (!)

Below in Section 2, the required form of the nonlinear transformations will
explicitly
be derived
and presented.
A short discussion of involved aspects will be given in the last Section 3.

Generally, when attempting to propose  a necessary definition of
{\it the scalar product}
$(X,Y)$ of two vectors related to the QH--numbers,
it is worth setting forth the conditions of
ordinary meaning and current
applicability. They should include:
\ses

\nin
$C_1$: the symmetry
\be
(X,Y)=(Y,X);
\ee
\ses

\nin
$C_2$: the normalization
\be
(X,X)^2=||X||;
\ee
\ses

\nin
$C_3$: the homogeneity
\be
(cX,Y)=(X,cY)=c(X,Y),
\ee
where
$c$ is any constant;
$cX$
means
the set
$\{cx^0,cx^1,cx^2,cx^3\}$
and
$cY$
means
the set
$\{cy^0,cy^1,cy^2,cy^3\}$;
\ses

\nin
$C_4$: the positivity
\be
(X,X)>0
\ee
(over all the sector
$V^{time--like}_4$);
\ses

\nin
$C_5$: the dilatation invariance
\be
(kX,kY)=(X,Y),
\ee
where
$kX$
and
$kY$
stay
for
$\{k^0x^0,k^1x^1,k^2x^2,k^3x^3\}$
and
$\{k^0y^0,k^1y^1,k^2y^2,k^3y^3\}$,
respectively.

Under these conditions, the required scalar product can be proposed as follows:
\be
\cS(X,Y)=\cS(A,B)
\ee
with
\be
\cS(A,B)\eqdef
\fr1{2F^2(B)}(A,B,B,B)
+
\fr1{2F^2(A)}(A,A,A,B),
\ee
where $A$ and $B$ relate to $X$ and $Y$, respectively;
\ses
\be
F(A)=\sqrt[4]{a^1a^2a^3a^4},
\qquad
F(B)=\sqrt[4]{b^1b^2b^3b^4},
\ee
such that
\be
\cS(A,A)=\lf(F(A)\rg)^2.
\ee

In an explicit coordinate way,
the definition
(1.7)--(1.8)
reads
\ses
$$
\cS(A,B)=
\fr1{8F^2(B)}(
a^1b^2b^3b^4+
a^2b^1b^3b^4+
a^3b^1b^2b^4+
a^4b^1b^2b^3)
$$
\ses
\ses
\be
+
\fr1{8F^2(A)}(
b^1a^2a^3a^4+
b^2a^1a^3a^4+
b^3a^1a^2a^4+
b^4a^1a^2a^3),
\ee
\ses\\
or
\ses
\be
\cS(A,B)=
\fr18
F(A)F(B)
\Bigl[
\fr{F(B)}{F(A)}
\lf(
\fr{a^1}{b^1}
+
\fr{a^2}{b^2}
+
\fr{a^3}{b^3}
+
\fr{a^4}{b^4}
\rg)
+
\fr{F(A)}{F(B)}
\lf(
\fr{b^1}{a^1}
+
\fr{b^2}{a^2}
+
\fr{b^3}{a^3}
+
\fr{b^4}{a^4}
\rg)
\Bigl].
\ee
All the conditions (1.1)--(1.5) are fulfilled.

In terms of the new variables
\be
d^1=\fr{b^1}{a^1},
\qquad
d^2=\fr{b^2}{a^2},
\qquad
d^3=\fr{b^3}{a^3},
\qquad
d^4=\fr{b^4}{a^4},
\ee
\ses\\
we get
\be
\cS=(F(A))^2F(D)\cL(D)
\ee
with
\ses
\be
F(D)=\sqrt[4]{d^1d^2d^3d^4}
\ee
\ses
and
\be
\ses
\cL(D)
\eqdef
\fr18
F(D)
\Bigl[
F(D)
\lf(
\fr{1}{d^1}
+
\fr{1}{d^2}
+
\fr{1}{d^3}
+
\fr{1}{d^4}
\rg)
+
\fr{1}{F(D)}
\lf(
d^1
+
d^2
+
d^3
+
d^4
\rg)
\Bigl].
\ee
On comparing
 the formulae (1.6)--(1.12) with the formulae
(1.13)--(1.15),
we may raise the conjecture  that
the invariance of the associated
function (1.15)
may entail the invariance of the initial scalar product (1.7).
In the next section, the conjecture will be confirmed by special
calculations.
\ses
\ses\\

\setcounter{sctn}{2}
\setcounter{equation}{0}

{\nin\bf 2. The Method of Calculations and The Main Result}
\ses

To our aims it is very convenient to adopt the parametrization
\be
d^1=\exp(
\de+\al+\beta+\ga
),
\ee
\ses
\be
d^2=\exp(
\de-\al+\beta-\ga
) ,
\ee
\ses
\be
d^3=
\exp(
\de+\al-\beta-\ga
)  ,
\ee
\ses
\be
d^4=
\exp(
\de-\al-\beta+\ga
),
\ee
where
$\{\al,\beta,\ga\}$
are the exponent variables.
We get
\be
F(D)=\exp(\de).
\ee
The inverse transition  reads as
\be
4\de=\ln d^1+\ln  d^2+\ln d^3+\ln  d^4,
\ee
\ses
\be
4\al=\ln d^1-\ln d^2+\ln d^3-\ln  d^4,
\ee
\ses
\be
4\beta=\ln d^1+\ln  d^2-\ln d^3-\ln  d^4,
\ee
\ses
\be
4\ga=\ln  d^1-\ln  d^2-\ln d^3+\ln  d^4.
\ee
\ses
In this way, the following representation
is obtained for the function
(1.15):
$$
\cL/F(D)=
\fr14\Bigl(
\cosh(\al+\beta+\ga)
+
\cosh(-\al+\beta-\ga)
+
$$
\ses
\be
\cosh(
\al-\beta-\ga)
+
\cosh(-\al-\beta+\ga)\Bigr).
\ee

The subsequent use of
the known hyperbolic identity
\be
\cosh\la-1=2(\sinh\fr{\la}2)^2
\ee
yields
$$
\cL/F(D)=
1+\fr12\Bigl[
(\sinh\fr{\al+\beta+\ga}2)^2
+
(\sinh\fr{-\al+\beta-\ga}2)^2
$$
\ses
\be
+(\sinh\fr{\al-\beta-\ga}2)^2
+(\sinh\fr{-\al-\beta+\ga}2)^2
\Bigr].
\ee
A careful consideration of the last function (2.12),
and of Eqs. (2.1)--(2.11),
suggests to apply the  transformations indicated below.
\ses
\ses\\

{\it
The $\{\al,\beta\}$--side turn}
acts as
\be
\de=\wt\de,
\qquad
\ga=\wt\ga,
\ee
\ses
and
\be
d^1=
e^{\wt\de+\wt\ga}
\exp
\Bigl[
2
\arcsinh
\Bigl(
\sinh\fr{\wt\al+\wt\beta}2\cos\mu
-
\sinh\fr{\wt\al-\wt\beta}2\sin\mu\Bigr)
\Bigr],
\ee
\ses
\ses
\be
d^2=
e^{\wt\de-\wt\ga}
\exp
\Bigl[
-2\arcsinh
\Bigl(
\sinh\fr{\wt\al-\wt\beta}2\cos\mu
+
\sinh\fr{\wt\al+\wt\beta}2\sin\mu\Bigr)
\Bigr],
\ee
\ses
\ses
\be
d^3=
e^{\wt\de-\wt\ga}
\exp
\Bigl[
2\arcsinh
\Bigl(
\sinh\fr{\wt\al-\wt\beta}2\cos\mu
+
\sinh\fr{\wt\al+\wt\beta}2\sin\mu\Bigr)
\Bigr],
\ee
\ses
\ses
\be
d^4=
e^{\wt\de+\wt\ga}
\exp
\Bigl[
-2
\arcsinh
\Bigl(
\sinh\fr{\wt\al+\wt\beta}2\cos\mu
-
\sinh\fr{\wt\al-\wt\beta}2\sin\mu\Bigr)
\Bigr].
\ee
\ses\\
By the help of the known
equality
$$
\arcsinh(x)=\ln\lf(x+\sqrt{x^2+1}\rg)
$$
we arrive at new convenient representations:
$$
d^1=
e^{\wt\de+\wt\ga}
\Biggl[
(
\sinh\fr{\wt\al+\wt\beta}2\cos\mu
-
\sinh\fr{\wt\al-\wt\beta}2\sin\mu
)
+\sqrt{(
\sinh\fr{\wt\al+\wt\beta}2\cos\mu
-
\sinh\fr{\wt\al-\wt\beta}2\sin\mu
)^2+1}
\,
\Biggr]^2,
$$
\ses
\ses
\ses
\ses
$$
d^2=
e^{\wt\de-\wt\ga}
\Biggl[
(
\sinh\fr{\wt\al-\wt\beta}2\cos\mu
+
\sinh\fr{\wt\al+\wt\beta}2\sin\mu
)
+\sqrt{(
\sinh\fr{\wt\al-\wt\beta}2\cos\mu
+
\sinh\fr{\wt\al+\wt\beta}2\sin\mu
)^2+1}
\,
\Biggr]^{-2},
$$
\ses
\ses
\ses
\ses
$$
d^3=
e^{\wt\de-\wt\ga}
\Biggl[
(
\sinh\fr{\wt\al-\wt\beta}2\cos\mu
+
\sinh\fr{\wt\al+\wt\beta}2\sin\mu
)
+\sqrt{(
\sinh\fr{\wt\al-\wt\beta}2\cos\mu
+
\sinh\fr{\wt\al+\wt\beta}2\sin\mu
)^2+1}
\,
\Biggr]^{2},
$$
\ses
\ses
\ses
\ses
$$
d^4=
e^{\wt\de+\wt\ga}
\Biggl[
(
\sinh\fr{\wt\al+\wt\beta}2\cos\mu
-
\sinh\fr{\wt\al-\wt\beta}2\sin\mu
)
+\sqrt{(
\sinh\fr{\wt\al+\wt\beta}2\cos\mu
-
\sinh\fr{\wt\al-\wt\beta}2\sin\mu
)^2+1}
\,
\Biggr]^{-2}.
$$

After that, we apply the pull-back substitutions
\be
4\wt\de=\ln\wt d^1+\ln \wt d^2+\ln \wt d^3+\ln\wt  d^4,
\ee
\ses
\be
4\wt\al=\ln\wt d^1-\ln \wt d^2+\ln \wt d^3-\ln\wt  d^4,
\ee
\ses
\be
4\wt\beta=\ln \wt d^1+\ln \wt d^2-\ln \wt d^3-\ln\wt  d^4,
\ee
\ses
\be
4\wt\ga=\ln \wt d^1-\ln\wt  d^2-\ln \wt d^3+\ln \wt d^4,
\ee
\ses
which entails
\be
\sinh\fr{
\wt\al+\wt\beta}2=
\fr12
\lf(
\sqrt[4]{
\fr{\wt d^1}{\wt d^4}}
-
\sqrt[4]{
\fr{\wt d^4}{\wt d^1}}
\rg)
,
\qquad
\sinh\fr{
\wt\beta-\wt\al}2=
\fr12
\lf(
\sqrt[4]{
\fr{\wt d^2}{\wt d^3}}
-
\sqrt[4]{
\fr{\wt d^3}{\wt d^2}}
\rg)
\ee
\ses
and
\be
e^{\wt\de+\wt\ga}
=
\sqrt{\wt d^1\wt d^4}
\ee
\ses
together with
\be
e^{\wt\de-\wt\ga}
=
\sqrt{\wt d^2\wt d^3}.
\ee

Thus we find
eventually the nonlinear turn given by
$$
d^1
(\mu;\wt d^1,\wt d^2,\wt d^3,\wt d^4)
=
\fr14
\sqrt{\wt d^1\wt d^4}
\,\Biggl[
\lf(
\sqrt[4]{
\fr{\wt d^1}{\wt d^4}}
-
\sqrt[4]{
\fr{\wt d^4}{\wt d^1}}
\rg)
\cos\mu
+
\lf(
\sqrt[4]{
\fr{\wt d^2}{\wt d^3}}
-
\sqrt[4]{
\fr{\wt d^3}{\wt d^2}}
\rg)
\sin\mu
$$
\ses
\ses
\be
+\sqrt{
\lf(
\lf(
\sqrt[4]{
\fr{\wt d^1}{\wt d^4}}
-
\sqrt[4]{
\fr{\wt d^4}{\wt d^1}}
\rg)
\cos\mu
+
\lf(
\sqrt[4]{
\fr{\wt d^2}{\wt d^3}}
-
\sqrt[4]{
\fr{\wt d^3}{\wt d^2}}
\rg)
\sin\mu
\rg)^2
+
4}
\,
\Biggr]^2,
\ee
\ses
\ses
\ses
\ses
\ses
\ses
\ses
\ses
$$
d^2
(\mu;\wt d^1,\wt d^2,\wt d^3,\wt d^4)
=
4
\sqrt{\wt d^2\wt d^3}
\Biggl[
-
\lf(
\sqrt[4]{
\fr{\wt d^2}{\wt d^3}}
-
\sqrt[4]{
\fr{\wt d^3}{\wt d^2}}
\rg)
\cos\mu
+
\lf(
\sqrt[4]{
\fr{\wt d^1}{\wt d^4}}
-
\sqrt[4]{
\fr{\wt d^4}{\wt d^1}}
\rg)
\sin\mu
$$
\ses
\ses
\be
+\sqrt{
\lf(
-
\lf(
\sqrt[4]{
\fr{\wt d^2}{\wt d^3}}
-
\sqrt[4]{
\fr{\wt d^3}{\wt d^2}}
\rg)
\cos\mu
+
\lf(
\sqrt[4]{
\fr{\wt d^1}{\wt d^4}}
-
\sqrt[4]{
\fr{\wt d^4}{\wt d^1}}
\rg)
\sin\mu
\rg)^2+4}
\,
\Biggr]^{-2},
\ee
\ses
\ses
\ses
\ses
\ses
\ses
\ses
\ses
$$
d^3
(\mu;\wt d^1,\wt d^2,\wt d^3,\wt d^4)
=
\fr14
\sqrt{\wt d^2\wt d^3}
\Biggl[
-
\lf(
\sqrt[4]{
\fr{\wt d^2}{\wt d^3}}
-
\sqrt[4]{
\fr{\wt d^3}{\wt d^2}}
\rg)
\cos\mu
+
\lf(
\sqrt[4]{
\fr{\wt d^1}{\wt d^4}}
-
\sqrt[4]{
\fr{\wt d^4}{\wt d^1}}
\rg)
\sin\mu
$$
\ses
\ses
\be
+\sqrt{
\lf(
-
\lf(
\sqrt[4]{
\fr{\wt d^2}{\wt d^3}}
-
\sqrt[4]{
\fr{\wt d^3}{\wt d^2}}
\rg)
\cos\mu
+
\lf(
\sqrt[4]{
\fr{\wt d^1}{\wt d^4}}
-
\sqrt[4]{
\fr{\wt d^4}{\wt d^1}}
\rg)
\sin\mu
\rg)^2+4}
\,
\Biggr]^{2},
\ee
\ses
\ses
\ses
\ses
\ses
\ses
\ses
\ses
$$
d^4
(\mu;\wt d^1,\wt d^2,\wt d^3,\wt d^4)
=
4
\sqrt{\wt d^1\wt d^4}
\,\Biggl[
\lf(
\sqrt[4]{
\fr{\wt d^1}{\wt d^4}}
-
\sqrt[4]{
\fr{\wt d^4}{\wt d^1}}
\rg)
\cos\mu
+
\lf(
\sqrt[4]{
\fr{\wt d^2}{\wt d^3}}
-
\sqrt[4]{
\fr{\wt d^3}{\wt d^2}}
\rg)
\sin\mu
$$
\ses
\ses
\be
+\sqrt{
\lf(
\lf(
\sqrt[4]{
\fr{\wt d^1}{\wt d^4}}
-
\sqrt[4]{
\fr{\wt d^4}{\wt d^1}}
\rg)
\cos\mu
+
\lf(
\sqrt[4]{
\fr{\wt d^2}{\wt d^3}}
-
\sqrt[4]{
\fr{\wt d^3}{\wt d^2}}
\rg)
\sin\mu
\rg)^2
+
4}
\,
\Biggr]^{-2}.
\ee

The nearest implications
\be
d^1d^2d^3d^4=
\wt d^1\wt d^2\wt d^3\wt d^4,
\ee
\ses
\ses
\ses
\be
d^1_{|_{\mu=0}}
=
\wt d^1,
\quad
d^2_{|_{\mu=0}}
=
\wt d^2,
\quad
d^3_{|_{\mu=0}}
=
\wt d^3,
\quad
d^4_{|_{\mu=0}}
=
\wt d^4,
\ee
\ses\\
and
\be
d^1d^4=
\wt d^1\wt d^4,
\qquad
d^2d^3=
\wt d^2\wt d^3
\ee
can readily be verified.

Also, on inserting (2.25)--(2.28) in  (1.15), we arrive at the invariance
\be
\cL(D)=\cL(\wt D)
\ee
after straightforward calculations.

Finally, we can return to  the initial variables in accordance with
\be
d^p=\fr{b^p}{a^p},
\qquad
\wt d^p=\fr{\wt b^p}{\wt a^p},
\ee
and
\be
a^p=\wt a^p,
\ee
 obtaining
$$
b^1
(\mu;\,a^1,a^2,a^3,a^4;\,\wt b^1,\wt b^2,\wt b^3,\wt b^4)
=
$$
\ses
\ses
$$
=
\sqrt{\fr{a^1}{a^4}}
\,\sqrt{\fr1{16}\wt b^1\wt b^4}
\,\Biggl[
\lf(
\sqrt[4]{
\fr{a^4\wt b^1}{a^1\wt b^4}}
-
\sqrt[4]{
\fr{a^1\wt b^4}{a^4\wt b^1}}
\rg)
\cos\mu
+
\lf(
\sqrt[4]{
\fr{a^3\wt b^2}{a^2\wt b^3}}
-
\sqrt[4]{
\fr{a^2\wt b^3}{a^3\wt b^2}}
\rg)
\sin\mu
$$
\ses
\ses
\ses
\ses
\ses
\ses
\be
+\sqrt{
\lf(
\lf(
\sqrt[4]{
\fr{a^4\wt b^1}{a^1\wt b^4}}
-
\sqrt[4]{
\fr{a^1\wt b^4}{a^4\wt b^1}}
\rg)
\cos\mu
+
\lf(
\sqrt[4]{
\fr{a^3\wt b^2}{a^2\wt b^3}}
-
\sqrt[4]{
\fr{a^2\wt b^3}{a^3\wt b^2}}
\rg)
\sin\mu
\rg)^2
+
4}
\,
\Biggr]^2,
\ee
\ses
\ses
\ses
\ses
\ses
\ses
\ses
\ses
\ses
\ses
$$
b^2
(\mu;\,a^1,a^2,a^3,a^4;\,\wt b^1,\wt b^2,\wt b^3,\wt b^4)
=
$$
\ses
\ses
$$
=
\sqrt{\fr{a^2}{a^3}}\,
\sqrt{16\wt b^2\wt b^3}
\Biggl[
-
\lf(
\sqrt[4]{
\fr{a^3\wt b^2}{a^2\wt b^3}}
-
\sqrt[4]{
\fr{a^2\wt b^3}{a^3\wt b^2}}
\rg)
\cos\mu
+
\lf(
\sqrt[4]{
\fr{a^4\wt b^1}{a^1\wt b^4}}
-
\sqrt[4]{
\fr{a^1\wt b^4}{a^4\wt b^1}}
\rg)
\sin\mu
$$
\ses
\ses
\ses
\ses
\ses
\ses
\be
+\sqrt{
\lf(
-
\lf(
\sqrt[4]{
\fr{a^3\wt b^2}{a^2\wt b^3}}
-
\sqrt[4]{
\fr{a^2\wt b^3}{a^3\wt b^2}}
\rg)
\cos\mu
+
\lf(
\sqrt[4]{
\fr{a^4\wt b^1}{a^1\wt b^4}}
-
\sqrt[4]{
\fr{a^1\wt b^4}{a^4\wt b^1}}
\rg)
\sin\mu
\rg)^2+4}
\,
\Biggr]^{-2},
\ee
\ses
\ses
\ses
\ses
\ses
$$
b^3
(\mu;\,a^1,a^2,a^3,a^4;\,\wt b^1,\wt b^2,\wt b^3,\wt b^4)
=
$$
\ses
\ses
$$
=
\sqrt{\fr{a^3}{a^2}}
\,
\sqrt{\fr1{16}\wt b^2\wt b^3}
\Biggl[
-
\lf(
\sqrt[4]{
\fr{a^3\wt b^2}{a^2\wt b^3}}
-
\sqrt[4]{
\fr{a^2\wt b^3}{a^3\wt b^2}}
\rg)
\cos\mu
+
\lf(
\sqrt[4]{
\fr{a^4\wt b^1}{a^1\wt b^4}}
-
\sqrt[4]{
\fr{a^1\wt b^4}{a^4\wt b^1}}
\rg)
\sin\mu
$$
\ses
\ses
\ses
\ses
\ses
\ses
\be
+\sqrt{
\lf(
-
\lf(
\sqrt[4]{
\fr{a^3\wt b^2}{a^2\wt b^3}}
-
\sqrt[4]{
\fr{a^2\wt b^3}{a^3\wt b^2}}
\rg)
\cos\mu
+
\lf(
\sqrt[4]{
\fr{a^4\wt b^1}{a^1\wt b^4}}
-
\sqrt[4]{
\fr{a^1\wt b^4}{a^4\wt b^1}}
\rg)
\sin\mu
\rg)^2+4}
\,
\Biggr]^{2},
\ee
\ses
\ses
\ses
\ses
\ses
\ses
\ses
\ses
\ses
\ses
$$
b^4
(\mu;\,a^1,a^2,a^3,a^4;\,\wt b^1,\wt b^2,\wt b^3,\wt b^4)
=
$$
\ses
\ses
$$
=
\sqrt{\fr{a^4}{a^1}}\,
\sqrt{16\wt b^1\wt b^4}
\,\Biggl[
\lf(
\sqrt[4]{
\fr{a^4\wt b^1}{a^1\wt b^4}}
-
\sqrt[4]{
\fr{a^1\wt b^4}{a^4\wt b^1}}
\rg)
\cos\mu
+
\lf(
\sqrt[4]{
\fr{a^3\wt b^2}{a^2\wt b^3}}
-
\sqrt[4]{
\fr{a^2\wt b^3}{a^3\wt b^2}}
\rg)
\sin\mu
$$
\ses
\ses
\ses
\ses
\ses
\ses
\be
+\sqrt{
\lf(
\lf(
\sqrt[4]{
\fr{a^4\wt b^1}{a^1\wt b^4}}
-
\sqrt[4]{
\fr{a^1\wt b^4}{a^4\wt b^1}}
\rg)
\cos\mu
+
\lf(
\sqrt[4]{
\fr{a^3\wt b^2}{a^2\wt b^3}}
-
\sqrt[4]{
\fr{a^2\wt b^3}{a^3\wt b^2}}
\rg)
\sin\mu
\rg)^2
+
4}
\,
\Biggr]^{-2}.
\ee

We have
\be
b^1b^2b^3b^4=
\wt b^1\wt b^2\wt b^3\wt b^4,
\ee
\ses
\ses
\ses
\be
b^1_{|_{\mu=0}}
=
\wt b^1,
\quad
b^2_{|_{\mu=0}}
=
\wt b^2,
\quad
b^3_{|_{\mu=0}}
=
\wt b^3,
\quad
b^4_{|_{\mu=0}}
=
\wt b^4,
\ee
\ses
and
\be
b^1b^4=
\wt b^1\wt b^4,
\qquad
b^2b^3=
\wt b^2\wt b^3.
\ee

Similarly to (2.32), the invariance
\be
\cS(A,B)=\cS(A,\wt B)
\ee
holds fine.
\ses
\ses\\

\setcounter{sctn}{3}
\setcounter{equation}{0}

{\nin\bf 3. Conclusions}
\ses

The
associative
commutative
Quadrahyperbolic  Numbers
come from several sources (see, e.g.,  [1--22]).
The fact is known, however,
 that the applied capacity of relativistic theories
is supported by their powerful invariance.
In the preceding section we have obtained the 1-parametric transformations
which are remarkable in that they act as nonlinear rotations around
a fixed vector
$A$
(see Eqs. (2.33)--(2.34)), leave invariant the respective scalar
product
(defined by Eq. (1.7)),
and simultaneously retain the length of rotated vectors $B$ unchanged
(as shown by Eq. (2.39)).
One can think of them  as defining
\ses

\nin
{\it The
$\{\al,\beta\}$-turn
of an
arbitrary vector
$B$
by the angle $\mu$ around a fixed vector  $A$.}
\ses

\nin
To show this, we have used
specific calculations in terms of associated exponents
(2.6)--(2.9).
Just similar calculations go through if
the $\{\al,\ga\}$--turn,
or
$\{\beta,\ga\}$--turn,
is dealt with.
\ses

The composition of these three transformations do
form a 3--parametrical group of
isotropy of the fixed vector $A$.
Whence
we have gained the new relativistic
framework in which the
6--parametrical
group of hypercomplex nonlinear
invariance should be substituted with
the
ordinary linear
pseudo-Euclidean
6--parametrical
Lorentz group.

It is hoped that the rigorous facts reported above
may favour due applications of
the
associative
commutative
 Quadrahyperbolic  Numbers
to various modern physical--relativistic subjects
that
 may be very different from what we are used to in
 bilinear (Euclidean
and pseudo--Euclidean)  theories.
\ses
\ses\\

\def\bibit[#1]#2\par{\rm\noindent\parskip1pt
                     \parbox[t]{.05\textwidth}{\mbox{}\hfill[#1]}\hfill
                     \parbox[t]{.925\textwidth}{\baselineskip11pt#2}\par}

\nin{\bf References}
\bigskip

 \bibit[1] D.G. Pavlov:
Hypercomplex Numbers, Associated Metric Spaces,
and Extension of
Relativistic Hyperboloid,
 gr-qc/0206004.

\bibit[1] H. Helmholtz:
\it \"{U}ber die Thatsachen  die der Geometrie zum
Grunde liegen, \rm G\"{o}ttingen, Nachr. 1868.

\bibit[2] I.L. Kantor, A.S. Solodovnikov:
{\it
Hypercomplex numbers: An
elementary introduction to algebras}, Springer, Berlin 1989.

\bibit[3] G. Birkhoff and S. MacLane: \it
Modern Algebra, \rm
Macmillan, New York, Third Edition
1965.

\bibit[4] B.L. van der Waerden:
\it Modern Algebra, \rm F.Ungar, New York, Third Edition 1950.

\bibit[5] O. Taussky:
\it
Algebra, \rm
\ in Handbook of Physics, ed. by E. U.Condon and H. Odishaw,
McGraw-Hill, New York, Second Edition 1958.

\bibit[6] J.W. Brown and R.V.Churchill:
\it Complex variables and
applications, \rm McGraw-Hill,  New York, 1996.

\bibit[7] R.L. Goodstein:
\it Complex functions, \rm McGraw-Hill,  New York,
1965.

\bibit[8] K. G\"{u}rlebeck and W.Spr\"{o}ssig:
\it Quaternionic and
Clifford calculus for physicists and engineers, \rm John Wiley\&Sons, 1997.

\bibit[9] G.B. Price:
\it
An introduction to multicomplex spaces and
functons, \rm
Marcel Dekker, 1991.

\bibit[10] J.B. Seaborn: \it
Hypergeometric functions and their applications,
\rm
Springer, Berlin 1991.

\bibit[11] A. van Proeyen:
Special geometries, from real to quaternionic,
hep-th/0110263.

\bibit[12] Y. Nutku, M.B. Sheftel:
A family of heavenly metrics,
gr-qc/0105088.

\bibit[13] G. Papadopoulos:
Brane Solitons and Hypercomplex Structures,
 math.DG/0003024.

\bibit[14] D.M.J. Calderbank
and P. Tod:
Einstein metrics, hypercomplex structures and the Toda field equation,
math.DG/9911121.

\bibit[15] M. Ansorg:
Differentially rotating disks of dust: Arbitrary rotation law,
 gr-qc/0006045.

 \bibit[16] S. Ollariu:
Complex numbers in n dimension,
 math.CV/9011077.

\bibit[17] D.M. J. Calderbank,
Einstein metrics, hypercomplex structures and the Toda field equation,
P. Tod, math.DG/9911121.

\bibit[18] D.L. Stefano:
Hypercomplex group theory,
 physics/9703033.

\bibit[21] S. R\"{o}nn:
Bicomplex algebra and function theory,
 math.CV/0101200.

\bibit[22] D.L. Stefano and R. Pietro:
Local Hypercomplex Analyticity,
funct-an/9703002.

\end{document}